\documentstyle[12pt]{article}
\def\Re{{\cal R \mskip-4mu \lower.1ex \hbox{\it e}\,}}
\def\Im{{\cal I \mskip-5mu \lower.1ex \hbox{\it m}\,}}
\def\ie{{\it i.e.}}
\def\eg{{\it e.g.}}

\def\etal{{\it et al.}}

\def\sub#1{_{\lower.25ex\hbox{$\scriptstyle#1$}}}
\def\sul#1{_{\kern-.1em#1}}
\def\sll#1{_{\kern-.2em#1}}
\def\sbl#1{_{\kern-.1em\lower.25ex\hbox{$\scriptstyle#1$}}}
\def\ssb#1{_{\lower.25ex\hbox{$\scriptscriptstyle#1$}}}
\def\sbb#1{_{\lower.4ex\hbox{$\scriptstyle#1$}}}

\def\slash{\not\!}

\def\to{\rightarrow}
\def\mh{\ifmmode m\sbl H \else $m\sbl H$\fi}
\def\mch{\ifmmode m_{H^\pm} \else $m_{H^\pm}$\fi}
\def\mt{\ifmmode m_t\else $m_t$\fi}
\def\mc{\ifmmode m_c\else $m_c$\fi}
\def\mz{\ifmmode M_Z\else $M_Z$\fi}
\def\mw{\ifmmode M_W\else $M_W$\fi}
\def\mws{\ifmmode M_W^2 \else $M_W^2$\fi}
\def\mhs{\ifmmode m_H^2 \else $m_H^2$\fi}
\def\mzs{\ifmmode M_Z^2 \else $M_Z^2$\fi}
\def\mts{\ifmmode m_t^2 \else $m_t^2$\fi}
\def\mcs{\ifmmode m_c^2 \else $m_c^2$\fi}
\def\mchs{\ifmmode m_{H^\pm}^2 \else $m_{H^\pm}^2$\fi}
\def\ztwo{\ifmmode Z_2\else $Z_2$\fi}
\def\zone{\ifmmode Z_1\else $Z_1$\fi}
\def\mtwo{\ifmmode M_2\else $M_2$\fi}
\def\mone{\ifmmode M_1\else $M_1$\fi}
\def\tb{\ifmmode \tan\beta \else $\tan\beta$\fi}
\def\xw{\ifmmode x\sub w\else $x\sub w$\fi}
\def\ch{\ifmmode H^\pm \else $H^\pm$\fi}
\def\lum{\ifmmode {\cal L}\else ${\cal L}$\fi}
\def\inpb{\ifmmode {\rm pb}^{-1}\else ${\rm pb}^{-1}$\fi}
\def\infb{\ifmmode {\rm fb}^{-1}\else ${\rm fb}^{-1}$\fi}
\def\epem{\ifmmode e^+e^-\else $e^+e^-$\fi}
\def\ppb{\ifmmode \bar pp\else $\bar pp$\fi}

\def\bsg{\ifmmode b\rightarrow s\gamma \else $b\rightarrow s\gamma$\fi}

\newskip\zatskip \zatskip=0pt plus0pt minus0pt
\def\matth{\mathsurround=0pt}

\def\atversim#1#2{\lower0.7ex\vbox{\baselineskip\zatskip\lineskip\zatskip
  \lineskiplimit 0pt\ialign{$\matth#1\hfil##\hfil$\crcr#2\crcr\sim\crcr}}}

\renewcommand{\thefootnote}{\fnsymbol{footnote}}

\begin{document} \begin{titlepage}
\setcounter{page}{1}
\thispagestyle{empty}
\rightline{\vbox{\halign{&#\hfil\cr
&SLAC-PUB-6475\cr
&April 1994\cr
&T/E\cr}}}
\vspace{0.8in}
\begin{center}

{\Large\bf
Single $W_R$ Production in $e^-e^-$ Collisions at the NLC}
\footnote{Work supported by the Department of
Energy, contract DE-AC03-76SF00515.}
\medskip

\normalsize THOMAS G. RIZZO
\\ \smallskip
{\it {Stanford Linear Accelerator Center\\Stanford University,
Stanford, CA 94309}}\\

\end{center}

\begin{abstract}

Single $W_R$ production in $e^-e^-$ collisions at the NLC can be used to
probe the Majorana nature of the heavy neutrinos present in the Left-Right
Symmetric Model below the kinematic threshold for their direct production.
For colliders in the $\sqrt {s}=1-1.5$ TeV range, typical cross sections of
order $1-10 fb$ are obtained, depending on the specific choice of model
parameters. Backgrounds arising from Standard Model processes are shown to be
small. This analysis greatly extends the kinematic range of previous studies
wherein the production of an on-shell, like-sign pair of $W_R$'s at the NLC
was considered.

\end{abstract}

\vskip0.45in
\begin{center}

Submitted to Physical Review {\bf D}.

\end{center}


\renewcommand{\thefootnote}{\arabic{footnote}} \end{titlepage}


One of the most attractive explanations for the apparently small magnitude of
neutrino masses is the see-saw mechanism{\cite {seesaw}} which can be easily
implemented within the framework of extended electroweak gauge theories(EEGT).
Such a scheme naturally leads to the prediction that the ordinary Standard
Model(SM) neutrinos are Majorana particles and that there must also exist
heavy, neutral, $SU(2)_L$-isosinglet fields ($N$) which are also Majorana
particles. Due to the Majorana nature of both sets of neutrinos, one expects
that the Lagrangian containing the corresponding mass terms can result in
the existence of new lepton-number violating, $\Delta L=2$ interactions at
a level that might be experimentally observable. At low energies, the best
example of such an interaction is neutrinoless double-$\beta$ decay
{\cite {double}} which has been sought for some time. In fact, existing limits
from searches for such decays must be used as input by model builders to
construct consistent scenarios for neutrino masses.  At higher energies, these
$\Delta L=2$ interactions can manifest themselves in many ways, \eg, once
produced, the $N$'s decay to charged leptons of both signs with equal rates.
One interesting possibility{\cite {old}}, that has recently been
revitalized{\cite {new}},
is the production of a like-charged pair of gauge bosons at an $e^-e^-$
collider. As has been much emphasized, high energy $e^-e^-$
collisions in the $0.5-1.5$ TeV range are a possible option at the
NLC{\cite {clem}}. As in ordinary neutrinoless double-$\beta$
decay($\beta\beta_{0\nu}$), this
reaction can {\it only} occur if the massive neutrinos
are Majorana particles, as lepton number is violated by two units. In some
sense, this reaction is really just the {\it inverse} process to
$\beta\beta_{0\nu}$.

The Left-Right Symmetric Model(LRM){\cite {lrm}}, which is based on the
extended weak gauge
group $SU(2)_L \times SU(2)_R \times U(1)$, provides a very natural setting for
the above scenario wherein the $N$'s are identified with heavy right-handed
neutrinos and $SU(2)_R$-breaking occurs via isotriplet scalars. In this model,
the cross section for $e^-e^- \to W_L^-W_L^-$ is
quite small (due to the fact that left-handed neutrino masses are tiny and
the various mixing angles are small) and so we are left considering the process
$e^-e^- \to W_R^-W_R^-$ as was done in Refs.{\cite {old,new}}. One difficulty,
however, is that it may not be possible to produce two, on-shell $W_R$'s at
the NLC with a center of mass energy in the range $0.5 \leq \sqrt{s} \leq 1.5$
TeV as $W_R$'s are generally expected to be rather heavy. It may, however, be
possible to examine the corresponding single-production process
$e^-e^- \to W_R^-(W_R^-)^* \to W_R^- +jj$ at the NLC, since it
only requires the $W_R$ to have a mass somewhat below $\sqrt {s}$ for this
final state to be kinematically accessible. If the resulting cross section
was found to be
sufficiently large, this would greatly extend the capability of the NLC to
probe the Majorana nature of the heavy neutrinos in this model as well as to
examine the interplay between
the $SU(2)_R$ symmetry breaking and Majorana mass generating mechanisms.
Of course, to produce only one on-shell $W_R$ we must pay the price of
the square of an additional electroweak gauge coupling, as well as
three-body phase space, both of which result in a corresponding
rate reduction in comparison to the
two-body $W_R^-W_R^-$ process. However, since typical $e^-e^- \to W_R^-W_R^-$
cross sections were found to be on the order of $1 pb$ or more, this
additional cost may not be too high if integrated luminosities in the
$100 fb^{-1}$ range are obtainable at the NLC for center of mass energies at or
above 1 TeV. In addition, this single-production process is expected to be
rather background-free since the on-shell $W_R$ can be easily reconstructed
from its decay products. It is the purpose of this paper to calculate the
cross section for this single-production process at NLC energies and determine
the corresponding event rates given the range of anticipated NLC luminosities.
Note that this process will allow us to probe the Majorana nature of $N$ even
though we may be below threshold for its direct production.

Constraints on the
mass of the $W_R$ ($M_R$) arise from many sources but are subject to various
different assumptions about the magnitudes of the undetermined parameters of
the LRM{\cite {lang}}: $\kappa=g_R/g_L$, the ratio of the $SU(2)_{L,R}$ gauge
couplings, the elements of the right-handed Cabibbo-Kobayashi-Maskawa mixing
matrix, $V_R$, and the masses for the right-handed neutrinos, $M_N$. The most
important of the existing constraints on $M_R$ are all severly weakened
if we are free to allow these unknown parameters to vary significantly from
such simplifying assumptions as $\kappa=1$, $V_R=V_L$, and that the $N$'s are
light. For example, the polarized $\mu$-decay bounds{\cite {mu}} (originating
from both $K$ and $\pi$ decays) are trivially avoided
if $M_N>50$ MeV. Likewise, the Tevatron bounds{\cite {tev}} can be easily
satisfied if {\it any} of the following are true: ($i$) the $N$'s decay in the
detector, ($ii$) if the $W_R$'s were to decay to many non-SM final states, or
($iii$) if $(V_R)_{ud}$ were much smaller than that suggested by the $V_R=V_L$
relationship. The well-known limit on $M_R$ from the $K_L-K_S$
mass difference{\cite {klks}} is also greatly weakened if $V_R=V_L$ is not
assumed and $\kappa <1$. (Grand unified models generally require{\cite {gut}}
that $\kappa \leq 1$ while consistency of the couplings in the Left-Right
Model requires $\kappa^2 \geq {x_w\over {1-x_w}}\simeq 0.303$, where
$x_w=sin^2 \theta_w$. Based on naturalness assumptions alone, we might expect
that $\kappa$ does not differ from unity by more than a factor of two.)
Putting all this together one finds that $M_R > 300$
GeV{\cite {me}}, in agreement with Ref. {\cite {lang}}, but we should
anticipate that $W_R$'s may be significantly heavier than this weak lower bound
would indicate, \ie, it is more than likely that if $W_R$'s do exist their
masses should be at least several times larger than this bound suggests.
Clearly, $W_R$'s more massive than about 700 GeV would be
too heavy to pair produce (on shell) at a $\sqrt {s}=1.5$ TeV collider with
a significant cross section; this forces us to consider the single production
scenario. We remind the reader that the mass of the $Z'$ in this model is
highly correlated with that of the $W_R$ and is a decreasing function of
$\kappa$; we find $M_{Z'}/M_R=3.55(1.69,~1.47)$ for $\kappa=0.6(1,~2)$.

An interesting case to consider is the possibility that a $W_R$ will not be
discovered until the NLC turns on, \ie, the LHC does not see $W_R$'s. This can
happen in the following way. Several analyses have shown{\cite {newwr}} that
$W_R$'s should be observable at the LHC in the mass range of our interest for
all reasonable values of $\kappa$ and the elements of $V_R$ {\it provided}
the decay $W_R \to eN$ is kinematically allowed. Essentially, the reason for
this is that both the machine and parton
luminosities are sufficient large (for $W_R$'s with masses of order 1 TeV) that
reduced effective couplings or leptonic branching fractions will not prevent
the $W_R$'s observation even if $N$'s are allowed to decay within
the detector volume.
However, it may be that $M_N > M_R$ so that the $W_R$ possesses {\it no}
leptonic decay modes. In this case searches in the dijet final state at the
LHC would be
necessary and, due to the tremendous QCD backgrounds, it is unlikely that a
$W_R$ could be discovered by such searches{\cite {wrjj}}. (Of course, a $W_R$
might be observable in the dijet mode {\it provided} its mass were already
known from measurements in other channels and sufficiently good mass
resolution was available.) If such a scenario
were realized, the NLC would play the role of discovery machine for $W_R$'s,
most likely in the $e^+e^-$ collider mode.

Unfortunately, to calculate the $e^-e^- \to W_R^-(W_R^-)^* \to W_R^- +jj$
cross section at the NLC, an additional free parameter is introduced in the
form of the doubly-charged Higgs scalar($\Delta$) mass, $M_{\Delta}$. The
reason for this is that the $W_R^-(W_R^-)^*$ state is produced by $N$-exchange
in the $t$- and $u$-channels together with a $\Delta$-exchange in
the $s$-channel.
All three contributions are required to maintain unitarity via gauge
cancellations as discussed in
Ref.{\cite {old,new}}. While the couplings of the $\Delta$ to both $e^-e^-$ and
$W_R^-W_R^-$ are fixed by the $SU(2)_R$ gauge symmetry breaking, the value of
$M_{\Delta}$ remains a free parameter in analogy to the Higgs boson mass
in the SM.
Thus the set of parameters we must consider when calculating the cross section
are $\kappa$, and $M_{N,R,\Delta}$. Fortunately, the cross section itself
scales with an overall factor of $\kappa^6$, which helps simplify our results.

Denoting the incoming $e^-$ momenta by $p_{1,2}$, the out-going, on-shell
$W_R$ momenta by $P$, and the final state fermion momenta by $q_{1,2}$, the
spin-summed, squared matrix element for
$e^-e^- \to W_R^-(W_R^-)^* \to W_R^- +f \bar f'$ is given by
\newpage
\begin{eqnarray}
|\bar {\cal M}|^2 & = & 4N_c\kappa^6 (g^2M_N/4)^2 ({g\over {2 \sqrt{2}}})^2
Tr(\slash {p_2}\Gamma_{\mu\nu}\slash {p_1}\Gamma_{\nu'\mu'}) (-g^{\mu\mu'}+
{P^{\mu}P^{\mu'}\over {M_R^2}})(g^{\nu\lambda}-{k^{\nu}k^{\lambda}
\over {M_R^2}}) \\ \nonumber
 & & (g^{\nu'\lambda'}-{k^{\nu'}k^{\lambda'}\over {M_R^2}})[(k^2-M_R^2)^2+
(\Gamma_RM_R)^2]^{-1}(q_{1\lambda}q_{2\lambda'}+q_{2\lambda}q_{1\lambda'}-
g_{\lambda\lambda'}q_1\cdot q_2) \,,
\end{eqnarray}
where $k=q_1+q_2$, $g=g_L$, $N_c$ is a color factor, and $\Gamma_R$ is
the width of $W_R$. Here, $\Gamma_{\mu \nu}$ is given by
\begin{eqnarray}
\Gamma_{\mu\nu} & = & {\gamma_\nu\gamma_\mu\over {t-M_N^2}}+
{\gamma_\mu\gamma_\nu\over {u-M_N^2}}+{4g_{\mu\nu}\over {s-M_{\Delta}^2}} \,,
\end{eqnarray}
with $s=(p_1+p_2)^2, ~t=(p_1-P)^2$, and $u=(p_2-P)^2$. Assuming massless final
state fermions, we integrate over their momenta and sum over all possible
flavour and color combinations leading to the differential cross section for
the $jj$ final state:
\begin{eqnarray}
{d\sigma\over {dz}} & = & 9{(G_FM_W^2/\pi)^3\over{24 \sqrt{2}}}\kappa^6
{M_N^2\over {s}}{\int_\delta^{1+\delta^2/4}}~dx~ \sqrt {x^2-\delta^2}
{1-x+\delta^2/4\over {(1-x)^2+(\Gamma_RM_R/s)^2}} R \,,
\end{eqnarray}
with $M_W$ being the SM $W$ mass, $z=cos \theta$,  and
($x,\delta)=2(E,M_R)/{\sqrt s}$, where $E$ is the energy of the on-shell
$W_R$. We define the angle $\theta$ to be that between the
three-vector components
of $P$ and $p_1$. It is important to note that the cross section is directly
proportional to $M_N^2$, thus it vanishes as the Majorana mass of the heavy
neutrino tends to zero. This is as expected since the reaction we're
considering is a $|\Delta L|=2$ process.
The expression for $R$ is rather complicated; let us first define the
following combinations of kinematic variables in order to simplify the
various contributions that appear below:
\begin{eqnarray}
k^2 & = & s+M_R^2-2E\sqrt {s} \,, \nonumber \\
\Sigma & = & s-M_R^2-k^2  \,, \nonumber \\
s_{\Delta} & = & s-M_{\Delta}^2 \,, \nonumber \\
t_{R,N} & = & t-M_{R,N}^2 \,, \nonumber \\
u_{R,N} & = & u-M_{R,N}^2 \,,  \\
G_{\Delta} & = & M_{\Delta}\Gamma_{\Delta} \,, \nonumber \\
f & = & (k^2M_R^2)^{-1} \,, \nonumber
\end{eqnarray}
with $\Gamma_{\Delta}$ being the width of the $\Delta$ which we obtain by
summing over the $e^-e^-$ and $W_R^-W_R^-$ decay modes.
In terms of the $W_R$ energy and scattering angle, the kinematic quantities
$t$ and $u$ are given by
\begin{eqnarray}
t & = & -\sqrt {s} E(1-\beta z)+M_R^2 \,, \nonumber \\
u & = & -\sqrt {s} E(1+\beta z)+M_R^2 \,,  \\
\beta & = & {\sqrt{x^2-\delta^2}\over {x}} \,. \nonumber
\end{eqnarray}
We now can write $R$ as
\begin{eqnarray}
R & = & a+b+c \,, \nonumber \\
a & = & 16s(s_{\Delta}^2+G_{\Delta}^2)^{-1}(2+f\Sigma^2/4) \,, \nonumber \\
b & = & 8s_{\Delta}(s_{\Delta}^2+G_{\Delta}^2)^{-1}[s(2+f\Sigma^2/4)(t_N^{-1}
+u_N^{-1})+f\Sigma(t_N^{-1}-u_N^{-1})(t_R^2-u_R^2)/4] \,, \nonumber \\
c_1 & = & s[(t_N^{-1}+u_N^{-1})^2+4t_N^{-1}u_N^{-1}] \,, \nonumber \\
c_2 & = & f(k^2+M_R^2)u_Rt_R(t_N^{-2}+u_N^{-2}) \,, \nonumber \\
c_3 & = & f\Sigma[(t_Rt_N^{-1})^2+(u_Ru_N^{-1})^2] \,, \\
c_4 & = & f\Sigma(s\Sigma-t_R^2-u_R^2)u_N^{-1}t_N^{-1} \,, \nonumber \\
c & = & c_1+c_2+c_3+c_4 \,. \nonumber
\end{eqnarray}
Here, `$a$' arises from the pure $s$-channel $\Delta$ exchange, `$c$' is the
summed contribution of both the $u$- and $t$-channel $N$ exchanges,
while `$b$' is
the interference between the the $s$- and $u,t$- channels.
As expected, the differential cross section is symmetric under the interchange
of $u$ and $t$. The angular distribution itself is generally found to be
quite flat owing to the rather large masses involved in the propagators and
the unitarity cancellation among the three exchanges. (This lack of sensitivity
to $z$ is found to be essentially independent of the choice of particle
masses so long as we restrict ourselves to parameter space regions that yield
large cross sections.) This
implies that mild acceptance cuts will not lead to any significant
alterations in the event rates we obtain below. This will be shown explicitly
after a brief discussion of the total cross section for $W_R +jj$ production.

Integrating over the $W_R$ production angle yields the total event rates
found in Figs.~1 and 2, in which we have set $\kappa=1$ and scaled by an
integrated luminosity of $100 fb^{-1}$. Fig.~1a shows the number of expected
$W_R+jj$ events, as a function of $M_R$, at a $\sqrt {s}=1$ TeV $e^-e^-$
collider for different choices of $M_N$ and $M_{\Delta}$. The results are seen
to be quite sensitive to the values of these mass parameters even when $M_R$
is fixed. In Fig.~1b(c), we fix $M_R=700$ GeV and plot the event rate
as a function of $M_N(M_{\Delta})$ for various values of $M_{\Delta}(M_N)$.
Typically, we see event rates of order several hundred/yr except near the
$\Delta$ resonance (where very large rates are obtained) or when $M_N$ is
small (as the cross section vanishes for massless $N$ since it probes the $N$'s
Majorana nature). Increasing $\sqrt {s}$ to 1.5 TeV, as shown in Fig.~2a, we
see substantial cross sections are obtainable even assuming $W_R$'s in
the 1-1.2 TeV mass
range for some parameter choices. Fixing $M_R=1$ TeV in Figs.~2b and c, we
again see reasonable event rates for most choices of $M_N$ and $M_{\Delta}$
assuming $\sqrt {s}=1.5$ TeV. The exact rate is, however, a sensitive probe of
both the $N$ and $\Delta$ masses. It is important to remember that the
$\Delta$ can appear as a resonance in the $e^-e^-$ channel.

To verify our claim that the $W_R$ angular distribution is quite flat for
choices of $M_N$ and $M_{\Delta}$ which yield large cross sections, we show
in Fig.~3 the angular distribution of a $W_R$ with a mass of 700 GeV
at at $\sqrt {s}=1$ TeV
$e^-e^-$ collider. For most choices of the input masses we obtain extremely
flat distributions, however, when $N$ is light a significant angular
dependence is observed. This is simply a result of the $t-$ and $u-$ channel
poles which develop as $M_N$ tends to zero. Of course, small $M_N$ also leads
to a small cross section, as shown in Figs.~1 and 2, as might be expected
since the matrix element vanishes in this massless limit. This substantiates
our claim above that when the cross section is large the corresponding angular
distribution is flat.

Potential backgrounds to the process $e^-e^- \to W_R^-(W_R^-)^* \to W_R^- +jj$
at the NLC are easily controlled and/or removed.
One might imagine, for example,
some contamination from the SM process $e^-e^- \to W_L^-W_L^- \nu \nu$, but
this can be easily eliminated by using missing energy cuts and demanding that
the $W_R$ final state be reconstructed from either the $jj$ or $eN \to eejj$
decay modes. (Since the on-shell $W_R$ decays to either $jj$ or $eN \to eejj$
there is no missing energy in the signal process.)
In addition, with polarized beams, we can take advantage of the
fact that $W_R$ couples via right-handed currents while any SM background must
arise only via left-handed currents. Within the LRM itself a possible
background could arise from a similar lepton-number conserving processes such
as  $e^-e^- \to W_R^-W_R^- NN$. The rate for such a process would be highly
suppressed since there are several additional powers of the weak coupling
and there is now a five-body final state for one virtual $W_R$. In
addition, the $N$'s are quite massive implying that such final states are most
likely to be kinematically inaccessible. Even if such a final state could be
produced, in comparison to the process we are
considering, the subsequent $N$ decays would lead to a final state with too
many
charged leptons and/or jets.

In this paper we have addressed the following points:

($i$) While $e^-e^- \to W_R^-W_R^-$ is an excellent probe of both the Majorana
nature of $N$ and the symmetry breaking sector of the Left-Right Symmetric
Model, it is more than likely that $W_R$'s are too massive to be pair
produced at the NLC if $\sqrt {s}=1-1.5$ TeV. This forces us to consider the
production of a single on-shell $W_R$ via the process
$e^-e^- \to W_R^-(W_R^-)^* \to W_R^- +jj$.

($ii$) Since the pair of on-shell $W_R$'s cross section was generally very
large, we would expect that the single $W_R$ rate would be significant if
integrated luminosities in the $100 fb^{-1}$ range were available. From the
explicit calculations we found that these expectations were indeed valid for
most of the model parameter space yielding cross sections of order
$1-10 fb^{-1}$.

($iii$) For values of the input parameters that lead to significant rates
the $W_R$ angular distribution was found to be rather flat implying that
angular cuts will not significantly reduce the cross sections. The rates
themselves were found to be quite sensitive to the particular values of the
masses of $N$ and $\Delta$. Masses for both these particles beyond the
kinematic reach of the NLC were found to be probed by the single $W_R$
production process.

($iv$) The process $e^-e^- \to W_R^-(W_R^-)^* \to W_R^- +jj$ can be used to
probe the Majorana nature of the heavy neutrinos present in the Left-Right
Symmetric Model even when they are too massive to be directly produced.

The NLC can provide an excellent probe into the detailed structure of the
Left-Right Symmetric Model.

\vskip.25in
\centerline{ACKNOWLEDGEMENTS}

The author would like to thank J.L.\ Hewett for discussions related to this
work. The author would also like to thank the
members of the Argonne National Laboratory High Energy Theory Group for use of
their computing facilities.

\newpage

%
\def\MPL #1 #2 #3 {Mod.~Phys.~Lett.~{\bf#1},\ #2 (#3)}
\def\NPB #1 #2 #3 {Nucl.~Phys.~{\bf#1},\ #2 (#3)}
\def\PLB #1 #2 #3 {Phys.~Lett.~{\bf#1},\ #2 (#3)}
\def\PR #1 #2 #3 {Phys.~Rep.~{\bf#1},\ #2 (#3)}
\def\PRD #1 #2 #3 {Phys.~Rev.~{\bf#1},\ #2 (#3)}
\def\PRL #1 #2 #3 {Phys.~Rev.~Lett.~{\bf#1},\ #2 (#3)}
\def\RMP #1 #2 #3 {Rev.~Mod.~Phys.~{\bf#1},\ #2 (#3)}
\def\ZP #1 #2 #3 {Z.~Phys.~{\bf#1},\ #2 (#3)}
\def\IJMP #1 #2 #3 {Int.~J.~Mod.~Phys.~{\bf#1},\ #2 (#3)}

\newpage

%
{\bf Figure Captions}
\begin{itemize}

\item[Figure 1.]{Event rates per $100 fb^{-1}$ for $W_R+jj$ production at a
1 TeV $e^-e^-$ collider assuming $\kappa=1$ (a) as a function of
$M_R$ for $M_N=M_{\Delta}=1$ TeV
(dots), $M_{\Delta}=1.2$ TeV and $M_N=0.4$ TeV (dashes), $M_{\Delta}=0.3$
and $M_N=0.1$ TeV (dash-dots), $M_{\Delta}=2,~M_N=0.6$ TeV (solid), or
$M_{\Delta}=1.8$ and $M_N=0.6$ TeV (square dots); (b) with $M_R=700$ GeV fixed
as a function of $M_N$ for $M_{\Delta}=0.3(0.6, 1.2, 1.5, 2)$ TeV
corresponding to the dotted(dashed, dash-dotted, solid, square-dotted) curve;
(c) as a function of $M_{\Delta}$ for $M_N=0.2(0.5, 0.8, 1.2, 1.5)$ TeV
corresponding to the dotted(dashed, dash-dotted, solid, square-dotted) curve.}
\item[Figure 2.]{Same as Fig.~1, but for a
1.5 TeV $e^-e^-$ collider. In (b) and (c), a $W_R$ mass of 1 TeV is assumed.}
\item[Figure 3.]{Angular distribution for a $W_R$ of mass 700 GeV produced at a
$\sqrt {s}=1$ TeV $e^-e^-$ collider for the same parameter choices as in
Fig.~1a.}
\end{itemize}

\end{document}